\documentclass[twocolumn,showpacs,pra]{revtex4}

\usepackage{graphicx,graphics,color,amsmath,stmaryrd,slashed}

\usepackage{times}

\begin{document}

\title{Two parameters scaling approach to Anderson localization of weekly interacting BEC}


\author{Jian Xu}
\author{Dan-Wei Zhang}
\author{Xin-Ding Zhang}
\author{Zheng-Yuan Xue\footnote{xuezhengyuan@yahoo.com.cn}}
\affiliation{Laboratory of Quantum Information Technology, and
School of Physics and Telecommunication Engineering, South China
Normal University, Guangzhou 510006, China}

\date{\today}

\begin{abstract}
We numerically study the Anderson localization of weekly interacting
Bose-Einstein condensate in a one-dimensional disordered potential.
We show that the interacting energy can not fully convert to the kinetic energy
and two parameters are needed to describe such
system completely, i.e., the density profile can be described with the
sum of two exponential functions. This is a new attempt for precise
description of systems with interplay of disorder and interaction.
\end{abstract}


\maketitle

\section{Introduction}

Disorder is ubiquitous in nature that  strongly affects the
properties of many physical systems even it is only a weak
perturbation. Fifty years ago, the localization of individual
particles or waves in a disordered crystal was  predicted by
Anderson \cite{P.W.Anderson}, and thus it is called Anderson
localization (AL). It can be understood as an effect of multiple
refection of a plane wave subject to random scattering or random
potential barriers \cite{N.F.Mott}. Later, AL has been unambiguously
observed in many systems under the approximation of a single
particle in a stationary disordered potential. For example,
electromagnetic waves in photonic lattices with disorder
\cite{T.Schwartz,Y.Lahini} and momentum distribution of quantum
kicked rotor
\cite{G.Casati1979,S.Fishman,G.Casati1989,F.L.Moore,J.Chabe}. In
single-particle approximation, the interaction among individual
particles is not taken into consideration. However, real materials
go far-beyond this approximation  \cite{Lee,B.Kramer}, and thus
observing AL is difficult. This can be understand as follows. Go
beyond single-particle approximation, one needs to consider the
interaction among particles. When this interaction induced nonlinear
effect presences, the reflected waves will interfere with each
other. Therefore, fully understanding the interplay of disorder and
interaction is an extremely difficult task both experimentally and
theoretically \cite{B.Kramer}.

Ultracold quantum gas possesses unprecedented possibility of
controlling almost all relevant physical parameters
\cite{L.Sanchez-Palencia2007,L.Sanchez-Palencia2008,A.Aspect,L.Sanchez-Palencia2010,G.Modugno,Zhu2011,bs},
and thus recognized as an ideal system for quantum simulation. A
particularly interesting aspect of the system is implementing random
speckle potential  using laser beams
\cite{R.Grimm,J.M.Huntley,P.Horak,D.Cl¨¦ment,Zhu2009}, which makes
this system suitable for observing AL \cite{bs}. Recently, two experimental groups have
reported the observation of localization of a noninteracting as well
as weak interacting Bose-Einstein condensate (BEC) in two different
kinds of disordered potentials \cite{G.Roati,J.Billy}. The final
state profile is  theoretically described by a single parameter
called localization length (LL)
\cite{P.Lugan,G.Kopidakis,A.S.Pikovsky,S.Flach,Ch.Skokos}. However, if disorder is switched on,  different from Piraud's work \cite{M. Piraud}, interacting energy of BEC does not fully convert to kinetic energy when it stop to expand in a disorder potential. Therefore, the approach that interacting energy converts into kinetic energy completely \cite{L.Sanchez-Palencia2007} does not applicable for the center of the localized
profile where interaction between particles can not be ignored.
Similar scenario is expected in disorder induced AL, where
one-parameter scaling theory is valid only for locally weak disorder \cite{sc}.
For strong disorder, the wavefunction is localized on just few sites
and after that a very small exponential tail follows \cite{scp}.
This fact naturally leads us to consider AL with two
parameters in interacting system.
To exclude the effect induced from disorder,
we consider the regime of week disorder, whcih one-parameter
description is valid for disorder induced AL.

In this paper, we consider a concrete example of one-dimensional
(1D) BEC with repulsive interaction in a random potential. We prove
that two parameters description of AL in such system is more
reasonable than that of single LL parameter. This is a new method to
describe the localized profile, which provides a new method for
studying AL with the interplay of disorder and interaction. The rest
of this paper is arranged as following. In section II, we describes
1D BEC system with repulsive interaction, and shows that there is
two different LL for the wing and center parts. In section III, as
the wing LL is well known, we present detail study of the center LL
focusing on its scaling law. In section IV, we give an approximate
analytic expression linking the density profile of atom to the two
LL and discuss the deviation of the expression. Finally, a brief
summary is given in section V.

\section{Two parameters description of AL}

Considering 1D BEC with repulsive interaction initially loaded
in a harmonic trapping potential
$V_{ho}(z)=\frac{1}{2}M\omega^2z^2$, where $M$ is the atomic mass
and $\omega$ is the trapping frequency. The effective 1D structure
can be achieved by applying an extremely tight harmonic vertical
confinement to froze the atomic motion in the other two dimensions.
The atomic interaction is effectively characterized by the $s$-wave
scattering with effective 1D interacting strength labeled as $g$,
which is  experimentally tunable using the Feshbach resonance
technique \cite{Chin}. Here, one considers BEC in weakly interacting
regime, i.e. $\overline{n}\gg M g/\hbar^2$, where $\overline{n}$ is
the average atomic density. Under the mean-field approximation, the
dynamics of the considered system is governed by the following
Gross-Pitaevskii (GP) equation
\begin{equation}
\label{GPEq} i\hbar\frac{\partial}{\partial t}\psi(z,t)
=\left[\frac{\hat{p}_z^2}{2M}+V_{ext}(z)+g|\psi(z,t) |^2
\right]\psi(z,t),
\end{equation}
where $\hat{p}_z$ is the momentum operator, $V_{ext}(z)$ is the whole external potential, and the
normalized wave function $\psi(z,t)$ is corresponding to a constant
total number of atoms $N=\int | \psi(z,t) |^2 dz$.

Now looking into the time-evolution of the BEC in a disordered
external potential governed by Eq. (\ref{GPEq}). To this end, the procedures
in the experiment \cite{J.Billy} is followed: Firstly,
prepare the 1D BEC at equilibrium of the harmonic potential without
disorder, i.e., $V_{ext}(z)=V_{ho}(z)$. In the Thomas-Fermi (TF)
regime ($\mu\gg \hbar\omega$), the initial wave function takes the
form of an inverted parabola \cite{L.Sanchez-Palencia2008}
\begin{equation}
\label{Ini_WF} \psi (z,0)=\sqrt{\left({\mu \over g} \right) \left(1-
{z^2 \over L_{TF}^2}\right)}\Theta \left(1-{|z|^2\over
L_{TF}^2}\right),
\end{equation}
where $\mu$ is the chemical
potential, $L_{TF}=\sqrt{2\mu/\omega^2}$ is the TF half-length and
$\Theta$ denotes the Heaviside step function. Secondly, at time
$t=0$, one switches off the harmonic potential and applies a
disorder potential along the expanding axis (i.e. $z$ axis) of the
BEC. It also assumes that the disorder potential  $V_d(z)$ is
generated by laser speckle method as in the experiment
\cite{J.Billy}. It is a random potential with a truncated negative
exponential single-point distribution \cite{J.W. Goodman}:
\begin{equation}
\label{P[V(z)]} P[V(z)]=\frac{\exp[-(V(z)+V_R)/V_R]}{V_R}\Theta \left(\frac{V(z)}{V_R}+1\right),
\end{equation}
 The average of disorder potential is set to be
$\langle V_d(z)\rangle =0$ and its correlation function
$C(z)=\langle V_d(z')V_d(z'+z)\rangle =V_R^2c(z/\sigma_R)$,
which $c(u)=\sin^2(u)/u^2$, $V_R=\sqrt{\langle V_d^2\rangle}$ is the
standard deviation and $\sigma_R$ is the correlation length.
Thus the external potential after $t=0$ is given by
$V_{ext}(z)=V_{d}(z)$. In the case of $\sigma_R/\xi_{int}<1$ that
$\xi_{int}=\sqrt{4 M \mu}$ is the initial healing length of BECs,
the density profile of the BEC will take a form of an
exponential-decay function when experiencing enough time \cite{J.Billy}.

To investigate the AL of the system, the time evolution
of the atomic density profile $n(z,t)=|\psi(z,t)|^2$ is worked out
\begin{equation}
\label{TimeEvolution} \psi(z,t)=\hat{T} \exp\left(
-\frac{i}{\hbar}\int_0^t H_{GP} dt\right) \psi(z,0),
\end{equation}
where the GP Hamiltonian is
$H_{GP}=\frac{\hat{p}^2_z}{2M}+V_d(z)+g|\psi(z,t) |^2$ and $\hat{T}$
is the time ordering operator. $\psi(z,t)$ is numerically calculated
by using the standard operator-split method. According
to Ref. \cite{Larson}, Eq. (\ref{TimeEvolution}) can be rewritten as
\begin{eqnarray}
\psi(z,t+\delta t)&=&\left\{\exp\left({-\frac{i\hat{p}^2_z}{4M\hbar}
\delta t}\right) \right.\notag\\
&&\times \exp\left\{{-\frac{i}{\hbar}
\left[V_d(z)+g|\psi(z,t)|^2\right] \delta
t}\right\}\\
&&\left.\times \exp\left({-\frac{i\hat{p}^2_z}{4M\hbar} \delta
t}\right)+\mathcal {O}({\delta t}^3)\right\}\psi(z,t),\notag
\end{eqnarray}
where the high-order term $\mathcal {O}({\delta t}^3)$ comes from
the non-commuting relation of the terms in $H_{GP}$. In the
sufficiently short time step $\delta t$, this term can be safely
neglected. Combining with the Fourier transform between the position
and momentum spaces, we can finally get the numerical solution of
$\psi(z,t)$ following the computation procedure step by step with
time step $\delta t$.

\begin{figure}[tbp]
\includegraphics[width=8cm]{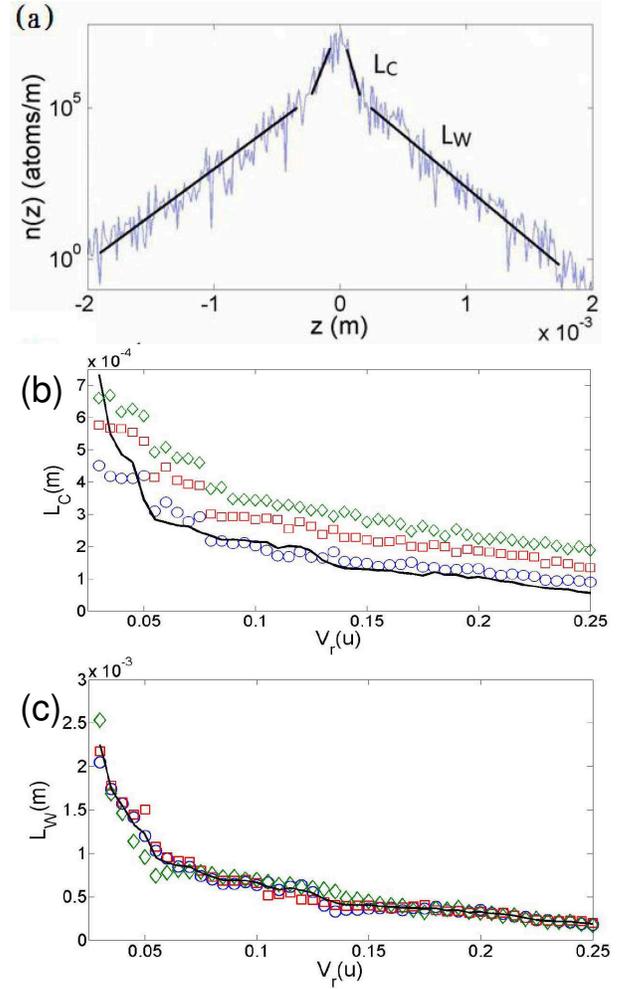}
\caption{(Color online) (a) The stationarity of the localized
profile with disorder potential strength $V_R=0.15\mu$,  nonlinear
intensity $Ng/h=0.035$ Hz, total number of atoms $N=1.7\times 10^4$
and $\sigma_R/\xi_{int}\approx 0.65$. We can define two
different LLs. Central LL (b) and  Wing LL (c) versus amplitude of the disordered
potential in different nonlinear intensity: $Ng/h$=0.035Hz (blue
$\times$), 0.104Hz (red $\nabla$), 0.204Hz (green $\circ $).
All LLs are obtained in an evolution time $t=2$ s. The black line indicate
the case of no interaction.}
\end{figure}

When experiencing long enough time, the BEC will stop expanding. As
$\sigma_R \ll L_{TF}$, the expanding of BEC is experienced with vast
random oscillation of potential. Due to disorder averaging, as
shown in Fig. 1(a), the final density profile of
BEC takes the form of  exponential-like function in a large
scale. It is find that the wing of density profile can be
exponentially fit by a wing LL denoted as $L_{W}$ in Fig. 1(a),
which has been considered as the {\sl single} parameter to
characterize LL of the system in experiment \cite{J.Billy} and takes
the analytic form of \cite{L.Sanchez-Palencia2007}
\begin{equation}
\label{LW} L_{W} =\frac{2\hbar^4}{\pi
M^2V_R^2\sigma_R\xi_{int}^2(1-\sigma_R/\xi_{int})}.
\end{equation}
However, the profile of the center part of BEC is not fit the form of
$L_{W}$ as shown in Fig. 1(a), which is also the case in Fig. (1) of the experiment
in Ref. \cite{J.Billy}.
In the experiment, the profile of center part of BEC is also stationary and do not fit the form of $L_{W}$. Further analysis
indicates that the result of this special center profile is that the interacting
energy does not convert kinetic energy completely, i.e., this profile
does not correspond to the standard AL.

In order to analyze it, one introduces a center LL denoted as $L_C$, which is
defined by the length in $z$ axis which reduces the maximum value of
the finial wave function by a factor of $1/e$.
When the interaction strength is weak ($Ng/h=0.035$ Hz), two LLs can be
approximately unified and the system may be described by the single parameter
theory \cite{L.Sanchez-Palencia2007}, which is verified by the very good
agreement of the black line and the green circle plot as shown in Fig. 1(b) and 1(c).
However, for stronger interaction ($Ng/h=0.104$ Hz),
both in experiment \cite{J.Billy} and our simulation in Fig. 1(b) indicate that
the central part of the density profile can not be
characterized by $L_{W}$ though it is also localized. To
describe the AL of the central part of the BEC, center LL is needed.

Furthermore, the trend of two LLs is studied by
changing the amplitude of the disordered potentials in different
nonlinear interaction intensity $Ng$. The results, as shown in Fig.1(b) and 1(c), show that the two LLs
both nearly exponentially decrease with the increasing of the
amplitude of the disordered potentials, but interestingly, they
exhibit different behavior: $L_{W}$ is insensitive to the nonlinear
intensity as expected from Eq. (\ref{LW}), but $L_{C}$ is strongly
affected by it. This can be understand by the facts: i) for larger
amplitude of the disordered potential, AL is more significant, and
thus both LLs are smaller; and ii) the impact of $g|\psi (z)|^2$ on
the wing part $L_W$ is slight, but significant for the
center part $L_C$ since the density of this part is much larger than
that of the wing part [cf. Fig. 1(a)]. It needs emphasizing that the extend of center part is much larger than $\sigma_R$, so it also undergo multiple refection random potential, which is AL characteristics.

\section{A scaling law of $L_C$}

In contrast to the single-parameter scaling theory for the disorder
system in liner regime, here two LLs as localization
parameters are needed to describe this nonlinear disorder system completely.
The difference in this nonlinear system is that  BEC finally reach
nearly equilibrium in the disorder potential, and then the
wing part can be well described by the LL for the noninteracting cases. However,  the central part still
contains certain residual interaction energy, and thus the LL of this part would be different.

Obviously, $L_C$ is relate to interaction. And numerically calculate finds that $L_C$ as a function of the
nonlinear interaction intensity $Ng$.
As shown in Fig. 2(a), one finds that $L_C$ approximately linearly
depends on $\sqrt{Ng}$, i.e.
\begin{equation}
\label{LC2}  L_{C}\propto \sqrt{Ng}+ constant,
\end{equation}
for a disorder potential with fixed $V_R$. For small $V_R$ cases,
there are some variances in the fitting (cf. Fig. 2(a)). Considering
the particle number $N$ with normalization equation
\begin{eqnarray}
\label{N}  N
&=&\int_{-\infty}^{\infty}|\psi(z,0)|^2dz=\int^{L_{TF}}_{-L_{TF}}
\frac{u}{g}\left(1-\frac{z^2}{L_{TF}^2}\right) dz\notag\\
&=&\frac{4\mu
L_{TF}}{3g},
\end{eqnarray}
one can obtain the expression of the TF half-length as
$L_{TF}=3Ng/4\mu$. It implies a relation
$L_C\propto\sqrt{L_{TF}}$. Combining with the dimensional analysis,
it is natural to guess that there may be a relationship between
$L_C$ and $\sqrt{L_{TF}}\times\sqrt{L_{W}}$.  Through a simple
fitting, It is find that $L_C$ has an interesting scaling law of
\begin{equation}
\label{LC3}  L_{C}\simeq\sqrt{L_{W} L_{TF}},
\end{equation}
as shown in Fig. 2(b).

\begin{figure}[tbp]
\includegraphics[width=8cm]{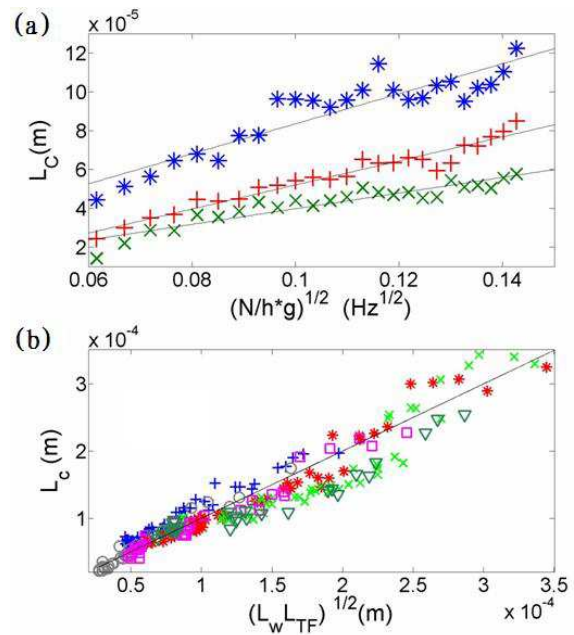}
\caption{(Color online) (a) The $L_{C}$ versus $\sqrt{g}$ in
different $V_R$: $V_R$=0.1u (blue $\ast$), 0.2u(red $+ $),
0.3u(green $\times$) with black lines are  theoretical predicted results
by Eq. (6). (b) The $L_{C}$ versus the $\sqrt{L_{W}L_{TF}}$. They are
obtained different nonlinear intensity: Ng/h=0.035Hz (blue
$\times$), 0.104Hz (red $\ast $), 0.204Hz (green +), and different
evolution time: 2.1s (gray $\circ $), 4.5s (purple $\oblong$), 7.0s
(dark green $\nabla$). The solid line is the function $y=x$.}
\end{figure}

It is showed some analysis about the physics picture of
Eq.(\ref{LC3}). $L_{W}$ can be taken as a main measurement of the
strength of localization of the system, thus $L_{C}$ as an
additional measurement should be positively related to $L_{W}$.
Since the central part of BEC can not fully expand and has less
kinetic energy, it is less sensitive to the strength of localization
than the wing part does. Therefore, its power index is less than $1$
and exhibits $L_{C}\propto \sqrt{L_{W}}$. Additionally, $L_{TF}$ is
a measurement of the size and the profile of the initial state, and
then $L_C$ could also be positively related to $L_{TF}$. Because of
the disorder potential and the smaller kinetic energy, the power
index is also less than $1$ and exhibits $L_{C}\propto
\sqrt{L_{TF}}$.

\section{An approximate analytic expression of density profile}

The previous analysis shows
that its finial density profile takes the form of
\begin{equation}
\label{nz1} n(z,\tau)\propto
 \begin{cases}
 \exp(-2|z|/L_{C})&\text{$|z|\leq z_0 $}\\
 \exp(-2|z|/L_{W})&\text{$|z|>z_0 $}
 \end{cases},
\end{equation}
where  $z_0$ is the theoretical cross-point. Here $z_0$ is
difficult to determine, so a simple form is assumed
\begin{equation}
\label{nz2}
n_a(z,\tau)=\frac{N-\Lambda}{L_{W}}\exp(-2|z|/L_{W})+\frac{\Lambda}{L_{C}}\exp(-2|z|/L_{C})
\end{equation}
to describe the full density profile, where $\Lambda$ is an
undetermined coefficient. In the following we will see that
$\Lambda$ can be approximately determined.

\begin{figure}[tbp]
\includegraphics[width=\columnwidth]{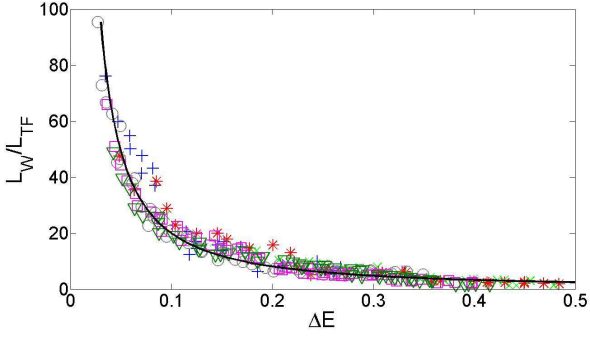}
\caption{(Color online) The scale parameter $L_{W}/L_{TF}$ versus
the residual interaction energy $\triangle E$ with $\sigma_R/\xi_{int}=0.65$.
$L_{W}/L_{TF}$ is obtained from different nonlinear intensity: 
0.104Hz (red  $\ast $), 0.204Hz (green +), and
different evolution time: 2.1s (gray $\circ$), 4.5s (purple
$\oblong$), 7.0s (dark green $\nabla $). The solid line is the best fit with the
function in the Eq.(10). }
\end{figure}

As the interaction energy does not transform
kinetic energy completely, considering  normalized residual interaction energy $\Delta
E$ to characterize this, which is defined as
\begin{eqnarray}
\label{deltaE1}
\Delta E &=&\frac{g\int |\psi(z,\tau)|^2dz}{g\int |\psi(z,0)|^2dz }\notag\\
 &=&\frac{5}{6N^2}
 \left[\Lambda^2\beta+(N-\Lambda)^2\beta^2+\frac{4(N-\Lambda)}{\beta^2+\beta}\right],
\end{eqnarray}
where $\tau$ is the typical time scale when the BEC is nearly
stable, and $\beta=\sqrt{L_{TF}/L_{W}}$. Considering Eq.(\ref{LC3}),
Eq.(\ref{deltaE1}) can become a simple form.
For the single LL cases, i.e. $\Lambda=0$ in Eq. (\ref{deltaE1}), It
can find that $\Delta E=5L_{TF}/6L_{W}$. Based on the numerical
simulation, It also finds that there is a well fitting formula to
characterize the relationship between $\Delta E$ and $L_w/L_{TF}$
for our two LLs description, as shown in Fig. 3. The fitting formula
is given by
\begin{equation}
\label{deltaE2}
 \frac{L_{W}}{L_{TF}}\simeq \triangle E^{-\frac{4}{3}}.
\end{equation}
This relation can be understood by the fact that $L_{W}$ and
$L_{TF}$ are the characteristic lengths of finial  and initial
states, and thus their ratio may has certain connection with the
residual interaction energy, which is the ratio of the interaction
energy of final and  initial states. If the BEC  expands fully, the
residual interaction energy tends to zero and $L_{W}$ will be
much larger than $L_{TF}$, corresponding to $\Delta E\rightarrow 0$
for $L_{W}/L_{TF}\rightarrow \infty$ in Eq. (\ref{deltaE2}) (cf.
Fig. 3). On the other hand, if the expansion is relatively very
small, comparable to the initial length scale of the BEC, then
$L_{W}$ will be nearly equivalent to $L_{TF}$, corresponding to the
residual interaction energy tends to unit.

Substituting Eq. (\ref{deltaE2}) into Eq. (\ref{deltaE1}), one can
work out $\Lambda$, which is approximately given by
\begin{equation}
\label{Lambda} \Lambda\simeq
\left[0.75\left(\frac{L_{TF}}{L_{W}}\right)^{0.1}+0.04\right]N
\end{equation}
Up to this, the full density profile of the
localized BEC can be characterized by Eq. (\ref{nz2}) and Eq.(\ref{Lambda}).

Due to Eq.(\ref{deltaE2}) and Eq.(\ref{Lambda}) are approximate,
it is need to investigate the deviation in the determining $L_C$ by using
Eq. (\ref{nz1}) and Eq. (\ref{nz2}). To this end, $z_1$
and $z_2$ ($z_1$, $z_2$$<z_0$) are definded as the solutions of
$n(z_1,\tau)=n_a(z_2,\tau)$ in the central part (i.e.
$z_{1,2}<L_C$):
\begin{equation}
\label{Eqn}
\left(\frac{N-\Lambda}{L_{W}}+\frac{\Lambda}{L_{C}}\right)e^{-2 \frac{z_1}{L_{C}}}
=\frac{N-\Lambda}{L_{W}}e^{-2 \frac{z_2}{L_{W}}}+\frac{\Lambda}{L_{C}}e^{-2 \frac{z_2}{L_{C}}}.
\end{equation}
and deviation
$\delta z=(z_2-z_1)/z_1$.
$\delta z>0$ indicate the distribution of our approximation is wider
than the precise distribution in the z-direction , and vice versa.
According to Eqs. (\ref{deltaE2}, \ref{Lambda}, \ref{Eqn}), the deviation $\delta z$ as a function of
$\Delta E$ and $z_1$ is showed in Fig. 4. And it can
find that when $\Delta E$ is in the small and large sides the
deviation is smaller than that in the intermediate regime. However,
the deviation using our approximation is always less than $18\%$ in
the whole regime. Relatively speaking, for non-interacting system,
both the deviation of the experimental and theoretical values are
larger than $50\%$  \cite{J.Billy}. While for the interacting
system, the deviation predicted in Ref.
\cite{L.Sanchez-Palencia2007} is larger than $60\%$ with the same
parameters of Fig. (1) when $V_R$ is in the small and large sides.
Therefore, It can conclude that Eq. (\ref{nz2}) of $n_a$ is a better
approximation to describe the density profile of the localized BEC
in a wide parameter range.

\begin{figure}[tbp]
\includegraphics[width=8cm,height=4.5cm]{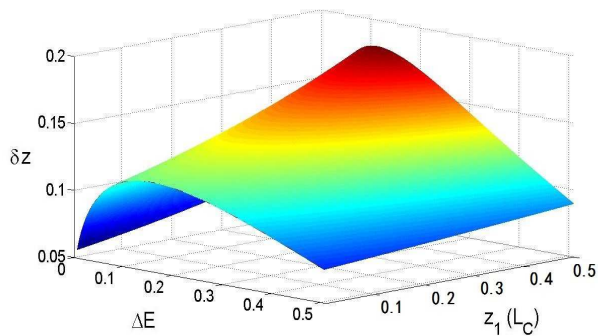}
\caption{(Color online) The parameter $\delta z$ versus the residual
interaction energy $\triangle E$ and $z_1/L_{C}$.}
\end{figure}

\section{Conclusion}

In summary, we have demonstrated that it is better and more
completely to use double LLs  to describe the AL of 1D weekly
interacting BECs in a disordered potential. We furthermore find a
scaling law related to the relationship between the newly defined LL
and the nonlinear atomic interactions. An approximate analytic form
of the full density profile of the localized BEC is also proposed by
using the two LLs.

\section*{Acknowledgements}
We thank Prof. Shi-Liang Zhu for  helpful discussions. This work was
supported by the NFRPC (No. 2013CB921804  and No. 2011CB922104), the NSFC (No. 11004065
and No. 10974059).

\end{document}